\documentclass[iop]{emulateapj}

\usepackage{graphicx}
\usepackage{amssymb}
\usepackage{amsmath}
\usepackage{natbib}
\usepackage{rotating}
\usepackage[FIGTOPCAP,tight]{subfigure}
\usepackage{hyperref}

\def\beqa{\begin{eqnarray}}
\def\eeqa{\end{eqnarray}}

\def\beq{\begin{equation}}
\def\eeq{\end{equation}}

\def\beqa{\begin{eqnarray}}
\def\eeqa{\end{eqnarray}}

\def\beq{\begin{equation}}
\def\eeq{\end{equation}}

\def\bseq{\begin{subequations}\begin{eqnarray}}
\def\eseq{\end{eqnarray}\end{subequations}}

\def\B{{\textsf{\textbf{B}}}}
\def\Bv{{\tilde{\B}}}

\def\u{{\bf{u}}}

\def\v{{\bf{v}}}

\def\I{{I}}

\def\Sv{{\textsf{\textbf{S}}}}

\def\sky{{\theta}}

\def\F{{\textsf{\textbf{F}}}}
\def\Ft{{\textsf{\textbf{F}}}^T}

\def\G{{\textsf{\textbf{G}}}}

\def\b{{\bf{b}}}

\def\G{{\textsf{\textbf{G}}}}

\def\H{{\textsf{\textbf{H}}}}

\def\UW{\altaffilmark{1}}
\def\UWtxt{\altaffiltext{1}{University of Washington, Seattle, USA}}

\def\Curtin{\altaffilmark{2}}
\def\Curtintxt{\altaffiltext{2}{International Centre for Radio Astronomy Research, Curtin University, Australia}}

\def\USwinburne{\altaffilmark{3}}
\def\USwinburnetxt{\altaffiltext{3}{Swinburne University of Technology, Melbourne, Australia}}

\def\CfA{\altaffilmark{4}}
\def\CfAtxt{\altaffiltext{4}{Harvard-Smithsonian Center for Astrophysics, Cambridge, USA}}

\def\ANU{\altaffilmark{5}}
\def\ANUtxt{\altaffiltext{5}{The Australian National University, Canberra, Australia}}

\def\ASU{\altaffilmark{6}}
\def\ASUtxt{\altaffiltext{6}{Arizona State University, Tempe, USA}}

\def\CSIRO{\altaffilmark{7}}
\def\CSIROtxt{\altaffiltext{7}{CSIRO Astronomy and Space Science, Australia}}

\def\Haystack{\altaffilmark{8}}
\def\Haystacktxt{\altaffiltext{8}{MIT Haystack Observatory, Westford, USA}}

\def\RRI{\altaffilmark{9}}
\def\RRItxt{\altaffiltext{9}{Raman Research Institute, Bangalore, India}}

\def\USydney{\altaffilmark{10}}
\def\USydneytxt{\altaffiltext{10}{University of Sydney, Sydney, Australia}}

\def\CAASTRO{\altaffilmark{11}}
\def\CAASTROtxt{\altaffiltext{11}{ARC Centre of Excellence for All-sky Astrophysics (CAASTRO)}}

\def\MIT{\altaffilmark{12}}
\def\MITtxt{\altaffiltext{12}{MIT Kavli Institute for Astrophysics and Space Research, Cambridge, USA}}

\def\UWNZ{\altaffilmark{13}}
\def\UWNZtxt{\altaffiltext{13}{Victoria University of Wellington, New Zealand}}

\def\UWisc{\altaffilmark{14}}
\def\UWisctxt{\altaffiltext{14}{University of Wisconsin--Milwaukee, Milwaukee, USA}}

\def\UMelbourne{\altaffilmark{15}}
\def\UMelbournetxt{\altaffiltext{15}{The University of Melbourne, Melbourne, Australia}}

\def\UTasmania{\altaffilmark{16}}
\def\UTasmaniatxt{\altaffiltext{16}{University of Tasmania, Hobart, Australia}}

\def\PerthUWA{\altaffilmark{17}}
\def\PerthUWAtxt{\altaffiltext{17}{Perth Observatory, Perth, Australia, and the University of Western Australia}}

\begin{document}

\title{Fast Holographic Deconvolution: a new technique for precision radio interferometry}

\author{
%Lead authors
I.~S.~Sullivan\UW,
M.~F.~Morales\UW,
B.~J.~Hazelton\UW,
%If separate collaboration author list, insert here
%Builders (comment out builders who appear above)
W.~Arcus\Curtin,
D.~Barnes\USwinburne,
G.~Bernardi\CfA,
F.~H.~Briggs\ANU,
J.~D.~Bowman\ASU,
%\newauthor
J.~D.~Bunton\CSIRO,
R.~J.~Cappallo\Haystack,
B.~E.~Corey\Haystack,
A.~Deshpande\RRI,
L.~deSouza\CSIRO$^,$\USydney,
D.~Emrich\Curtin,
B.~M.~Gaensler\USydney$^,$\CAASTRO,
%\newauthor
R.~Goeke\MIT,
L.~J.~Greenhill\CfA,
D.~Herne\Curtin,
J.~N.~Hewitt\MIT,
M.~Johnston-Hollitt\UWNZ,
D.~L.~Kaplan\UWisc,
J.~C.~Kasper\CfA,
%\newauthor
B.~B.~Kincaid\Haystack,
R.~Koenig\CSIRO,
E.~Kratzenberg\Haystack,
C.~J.~Lonsdale\Haystack,
M.~J.~Lynch\Curtin,
S.~R.~McWhirter\Haystack,
%\newauthor
D.~A.~Mitchell\CAASTRO,
E.~Morgan\MIT,
D.~Oberoi\Haystack,
S.~M.~Ord\CfA,
J.~Pathikulangara\CSIRO,
%\newauthor
T.~Prabu\RRI,
R.~A.~Remillard\MIT,
A.~E.~E.~Rogers\Haystack,
A.~Roshi\RRI,
J.~E.~Salah\Haystack,
%\newauthor
R.~J.~Sault\UMelbourne,
N.~Udaya~Shankar\RRI,
K.~S.~Srivani\RRI,
J.~Stevens\CSIRO$^,$\UTasmania,
R.~Subrahmanyan\RRI,
%\newauthor
S.~J.~Tingay\Curtin$^,$\CAASTRO,
R.~B.~Wayth\Curtin,
M.~Waterson\Curtin,
R.~L.~Webster\CAASTRO$^,$\UMelbourne,
%\newauthor
A.~R.~Whitney\Haystack,
A.~Williams\PerthUWA,
C.~L.~Williams\MIT,
J.~S.~B.~Wyithe\CAASTRO$^,$\UMelbourne
}

%Institutional footnotes (typeset, then rearrange here to be in order)
\UWtxt
\Curtintxt
\USwinburnetxt
\CfAtxt
\ANUtxt
\ASUtxt
\CSIROtxt
\Haystacktxt
\RRItxt
\USydneytxt
\CAASTROtxt
\MITtxt
\UWNZtxt
\UWisctxt
\UMelbournetxt
\UTasmaniatxt
\PerthUWAtxt

\begin{abstract}
We introduce the Fast Holographic Deconvolution method for analyzing interferometric radio data. Our new method is an extension of A-projection/software-holography/forward modeling analysis techniques and shares their precision deconvolution and widefield polarimetry, while being significantly faster than current implementations that use full direction-dependent antenna gains. Using data from the MWA 32 antenna prototype, we demonstrate the effectiveness and precision of our new algorithm. Fast Holographic Deconvolution may be particularly important for upcoming 21~cm cosmology observations of the Epoch of Reionization and Dark Energy where foreground subtraction is intimately related to the precision of the data reduction.
%This new method implements two important changes from previous approaches:  it subtracts the model from the holographic dirty images instead of the visibilities, and it introduces the sparse Holographic Mapping Function that increases the speed of holographic deconvolution by three orders of magnitude. 
\end{abstract}

\maketitle

\section{Introduction}
\setcounter{footnote}{0}

Interferometric radio arrays are some of the most scientifically productive tools of astrophysics. In particular, by choosing the appropriate antenna configuration the angular sensitivity of an interferometer can be tuned to the scientific question at hand. However, this flexibility comes at a the cost of %a very complicated point spread function (PSF, or array beam) and 
increased demands on the precision of the data analysis. 

%The CLEAN algorithm has been used for decades with very good success, but in the last few years there have been a number of important advances in the area of interferometric deconvolution including \citet{Bhatnagar:2008p3407}, \cite{Morales:2009p4730} \cite{Myers:2003p4911}, \citet{Carozzi:2009p3904} \citet{Bernardi:2011p4917}, \citet{Pindor:2011p4919}, and \citep{Li:2011p4921} to name just a few. In this article we present a new algorithm we call Fast Holographic Deconvolution (FHD) that includes the direction dependent gain and polarization correction of A-projection/software-holography/peeling, while being significantly faster than current approaches.

Radio interferometry uses an array of antennas to sample the electric field from cosmic sources. Each antenna collects the electric field across its surface, which is then cross-correlated against all of the other antenna measurements to generate pairwise visibilities. Each visibility is a measurement of an angular Fourier mode of the sky (for an introduction to interferometry see \citealt{Rohlfs}). The sparse sampling of angular Fourier modes allows a unique combination resolution and angular sensitivity tailored to the science case, at the expense of a complicated point spread function. Additionally, each antenna integrates the electric field across the antenna surface---fundamentally erasing information below the antenna scale and imprinting a direction-dependent gain and polarization response to the signal (see chapter 3 of \citealt{Morales:2010p4786} for a conceptual introduction to direction-dependent antenna effects). Recent advances by \cite{Bhatnagar:2008p3407}, \cite{Morales:2009p4730}, and \cite{Myers:2003p4911}---variously referred to as ``A-projection'' or ``software holography''---have enabled these direction-dependent gain and polarization effects to be included in the deconvolution process. While the resulting analyses are quite precise, they are slow compared to traditional CLEAN methods \citep{Hogbom1974}, particularly if all the antennas do not share the same gain pattern.

% (array beam) arising from the uneven sampling of modes creates a challenge for analysis algorithms. Precision deconvolution of both the point spread function and the individual antenna beam patterns from the true sky signal therefore lie at the heart of interferometric imaging.

%Recent advances in interferometric deconvolution have focused on including the direction-dependent sensitivity (gain) and polarization of each antenna. Variously referred to as A-projection, software holography, and peeling, these new techniques use the measured holographic antenna response to include direction-dependent gain and polarization effects in the deconvolution process. While the resulting analyses are quite precise \citep{Bhatnagar:2008p3407}, they are slow compared to traditional CLEAN methods \citep{Hogbom1974}, particularly if the antennas do not share the same gain pattern. 

In this article we introduce the Fast Holographic Deconvolution technique which increases the speed of A-projection/software-holography deconvolution. In \S \ref{sec:Alg} we introduce the Fast Holographic Deconvolution algorithm and compare it to commonly used radio deconvolution algorithms. We then describe our particular implementation of Fast Holographic Deconvolution in \S \ref{sec:FHD}, and apply the algorithm to data from the 32 antenna MWA prototype in \S \ref{sec:Results}. We conclude with a discussion of future work in \S \ref{sec:Conc}.

%Many techniques greatly expand the $u,v$ coverage of an instrument, such as sky rotation and co-adding fine frequency channels, but even then the coverage will have uneven weighting. Additionally, any realistic antenna has a non-negligble length, which means that every measurement is actually an integration over the area spanned by the antenna, and is not simply a precise measurement at a point. As a result, the beam pattern of the antenna is intrinsically imprinted in the data, and a straightforward discrete Fourier transform of the visibility data will not be strictly accurate. Thus, it is necessary to deconvolve both the antenna beam pattern\footnote{The XXX beam} and the coverage and weighting of the $u,v$-plane from the antenna distribution pattern\footnote{The primary beam}, neither of which can be done analytically. In this paper we introduce Fast Holographic Mapping, which is a new technique to use with well-established deconvolution algorithms that correctly incorporate both effects, but in a fraction of the time.

\section{Algorithm} 
\label{sec:Alg}
%In this section we briefly review these two styles of deconvolution and introduce the Fast Holographic Deconvolution algorithm. While this overly broad classification does not do justice to the many important advances in deconvolution, Fast Holographic Deconvolution is distinct in its approach from most common algorithms and this simplification helps place our new method in context.

There are many variations of radio deconvolution in the literature, including CLEAN \citep{Hogbom1974}, Cotton-Schwab faceted CLEAN \citep{Schwab1984,Cotton2004,Cotton2005}, multi-scale CLEAN, MEMs, A-projection based CLEAN \citep{Bhatnagar:2008p3407,Carozzi:2009p3904}, peeling \citep{Mitchell:2008p3967}, forward-modelling \citep{Bernardi:2011p4917,Pindor:2011p4919}, measurement equation \citep{Smirnov:2011p4926}, and compressive sampling \citep{Li:2011p4921}, to name just a few. Very crudely, the most popular styles of deconvolution can be divided into two families:  basic CLEAN implementations that subtract model visibilities from the data, and peeling, A-projection, and forward modeling approaches that subtract improved visibility estimates from the data with direction-dependent antenna calibration. While this overly broad classification does not do justice to the many important advances in deconvolution, this simplification helps place the advances of Fast Holographic Deconvolution in context. 

%In comparing the two broad styles of deconvolution, the variants of CLEAN have the advantage that they are very fast, though they are limited by being able to deconvolve source components only at pixel centers. As a result, sources must be fit as a collection of positive and negative components at discrete locations, which fundamentally limits precision. Most variants of CLEAN are also restricted to a narrow field of view, though techniques such as the Cotton-Schwab CLEAN \citep{Schwab1984} can incorporate wide-field effects at the expense of speed. In contrast, peeling and A-projection styles of deconvolution do not share these limitations and show promise for achieving very high dynamic range in wide-field images, but they are limited by being extremely computationally expensive. 

\subsection{CLEAN}
\label{sec:AlgCLEAN}
%\vspace{10 pt}
%\noindent \textbf{CLEAN}

The basic CLEAN algorithm loop---omitting for clarity widefield w-projection \citep{Cornwell:2008p4785}, faceting, major-minor cycles and other innovations---can be approximated with the following equation to determine the next component of the model:
\beqa
\gamma\,{\rm Max}\Bigg\{\Ft(\sky,\u)\Sv(\u,\v) \hspace{2 cm} \nonumber \\
\left(\G(\v,\v)\v_d - \sum_m \v_m \right)\Bigg\},
\label{CLEANeq}
\eeqa
where we use the linear algebra notation ${\textsf{\textbf{A}}}({\bf a},\b; x)$ to signify a transformation from coordinate vector $\b$ to coordinate vector ${\bf a}$ that depends on parameters $x$, and vector $\u$ is the $uv$ plane and $\v$ represents a list of visibilities.
Describing Equation \ref{CLEANeq} in words, the data visibilities $\v_d$ are calibrated ($\G$) and the model visibilities $\v_m$ for all the previously found components are subtracted. The terms in parentheses are then the residual visibilities which are gridded to the $uv$ plane ($\Sv$) and Fourier Transformed ($\F^T$) to form the residual dirty map. The peak of the map is selected and the loop gain $\gamma$ is applied to determine the next component of the model and the loop is repeated.

The model visibilities $\v_m$ are calculated by Fourier transforming the model source components $\I(\sky_{m})$ to the $uv$ plane and taking the $\delta$-function of the correlation function at the baseline separations, as shown in this annotated and expanded version
%\beq
%\gamma\,{\rm Max}\left\{\Ft(\sky,\u)\Sv(\u,\v)\left( \G^T(\v,\v)\v_d -  \sum_m \delta(\v,\u)\F(\u,\sky)\I(\sky_{m'})\right)\right\}.
%\eeq
%annotated
\beqa
\gamma\,{\rm Max}\Bigg\{\Ft(\sky,\u)\overbrace{\Sv(\u,\v)}^\textrm{gridding} \Bigg( \overbrace{\G(\v,\v)\v_d}^\textrm{calibrated data} - \nonumber \\ \overbrace{\sum_m \delta(\v,\u)\F(\u,\sky)\I(\sky_{m})}^\textrm{model visibilities} \Bigg)\Bigg\}. 
%\I(\sky_{\rm next}) = \gamma\,{\rm Max}\left\{\Ft(\sky,\u)\Sv(\u,\v)\left( \overbrace{\G^T(\v,\v)\v_d}^\textrm{calibrated data} -  \overbrace{\sum_m \underbrace{\delta(\v,\u)}_\textrm{baseline center}\F(\u,\sky)\underbrace{\I(\sky_{m'}}_\textrm{model sources})}^\textrm{model visibilities} \right)\right\}
\eeqa

This is of course a gross simplification of the algorithms in common use, but it captures the three key features of most CLEAN implementations:  the model visibilities are the component correlations at the baseline separation, source components are at pixel centers with a magnitude determined by a CLEAN gain $\gamma$ that is less than one, and the model is subtracted from the data as visibilities.

\subsection{A-projection}
\label{sec:AlgAproj}
%\vspace{10 pt}
%\noindent \textbf{A-projection, Peeling, and Forward Modeling }

%Peeling and A-projection make three significant changes to the basic CLEAN algorithm. First they use the software-holography/A-projection method of gridding ($\Bv$) where the holographic antenna beam is used to grid residual visibilities to the $uv$ plane. Second the beam pattern ($\B$) is used to calculate the model visibilities, and thirdly the model component is centroided instead of being at the center of the brightest pixel. 

A-projection makes two significant changes to the basic CLEAN algorithm \citep{Bhatnagar:2008p3407}. A-projection uses the holographic beam pattern ($\Bv$)  to grid residual visibilities to the $uv$ plane, effectively applying direction-dependent calibration, and it uses the beam pattern ($\B$) to calculate more accurate model visibilities. 
\beqa
\gamma\,{\rm Max}\Bigg\{\Ft(\sky,\u)\Bv^T(\u,\v) \hspace{1 cm} \nonumber \\ 
\Bigg(\v_d - \sum_m \B(\v,\u)\F(\u,\sky)\I(\sky_{m})\Bigg)\Bigg\}.
\label{PeelingEq}
\eeqa
%annotated
%\beq
%\I(\sky_{\rm next}) = \gamma\,{\rm MaxCentroid}\left\{\overbrace{\Ft(\sky,\u)\Bv^T(\u,\v)}^\textrm{RTS pipeline}\left(\v_d -  \sum_m \overbrace{\Bv(\v,\u)\F(\u,\sky)}^\textrm{MAPS}\I(\sky_{m})\right)\right\}
%\eeq
In words, the model components $\I(\sky_{m})$ are transformed to the $uv$ plane which is integrated by the known antenna response $\B(\v,\u)$ to form an improved estimate of the visibilities. Many array simulators such as MAPS (MIT Array Performance Simulator\footnote{\url{http://www.haystack.mit.edu/ast/arrays/maps/index.html}}) operate by performing the $uv$ integration described by $\B(\v,\u)$. This improved model visibility is subtracted from the measured data, and the residual visibilities are gridded using the A-projection/software holography kernels $\Bv$ to form a  holographic dirty map.
%\footnote{One confusing subtlety is that the holographic dirty map is in different units than a traditional dirty map. A traditional dirty map is in apparent magnitude---source brightness decreases away from field center with the average antenna beam---while in holographic dirty maps source brightness falls with the \textit{square} of the average antenna beam.} 
The peak of the image is used to produce the next model component, and the process is repeated.

In some situations it is faster to apply direction dependent gains ($\G(\v,\v;a,\sky_m)$) to each model visibility rather than integrate the $uv$ plane with $\B(\v,\u)$, giving
\beqa
\gamma\,{\rm MaxCentroid}\Bigg\{\Ft(\sky,\u)\Bv^T(\u,\v) \hspace{2.5 cm} \nonumber \\ 
\Bigg(\v_d -  \sum_m \G^T(\v,\v;a,\sky_m)\delta(\v,\u)\F(\u,\sky)\I(\sky_{m})\Bigg)\Bigg\}.
\label{PeelingEq2}
\eeqa
Mathematically these are identical as $\B(\v,\u)= \G(\v,\v;a,\sky_m)\delta(\v,\u)$, but they can have very different speeds depending on context. Using the direction dependent gain is sometimes referred to as peeling \citep{Mitchell:2008p3967}. This is also similar to the measurement equation approach to direction-dependent calibration developed by \citet{Smirnov:2011p4925,Smirnov:2011p4924}.

\subsection{Forward Modelling}
Forward modeling \citep{Bernardi:2011p4917,Pindor:2011p4919} is mathematically similar to A-projection and peeling, but subtracts holographic model and data images instead of the model and data visibilities:
\beqa
\gamma\,{\rm MaxCentroid}\Bigg\{ \left[\Ft(\sky,\u)\Bv^T(\u,\v)\v_d\right]_1 - \hspace{1 cm} \nonumber \\ 
\Ft(\sky,\u)\Bv^T(\u,\v)\B(\v,\u) \sum_m \F(\u,\sky)\I(\sky_{m})\Bigg\}.
\label{ImagingFHDEq}
\eeqa
The data is gridded and imaged [shown in square brackets] to form a holographic image (first line), and the model is converted to visibilities then gridded and imaged to form a holographic model image (second line), which are then subtracted. For traditional dirty images image subtraction would cause a significant degradation in deconvolution fidelity. However, because the antenna beam pattern ($\Bv$) is used to grid the data, holographic dirty maps contain all the information that was in the visibilities and   subtracting images or  visibilities works equally well (assuming calibration $\Bv$ is correct, \citealt{Morales:2009p4730}). This lossless information property was proved by \cite{Tegmark:1997p2009} and is widely used for reducing CMB data. One advantage of subtracting images is that the data only needs to be gridded and imaged once [square brakets]. Forward modeling also usually centroids the position of the source components as opposed to using the pixel centers, though component centroiding can be added to any of the deconvolution approaches.

\subsection{Fast Holographic Deconvolution}
\label{sec:AlgFHD}
%\vspace{10 pt}
%\noindent \textbf{Fast Holographic Deconvolution}
 \nopagebreak
 
The major change behind Fast Holographic Deconvolution is to introduce the Holographic Mapping function $\H(\u,\u)\equiv\Bv^T(\u,\v)\B(\v,\u)$. In all of the holographic deconvolution algorithms (A-projection Eq. \ref{PeelingEq}, peeling with holographic gridding Eq. \ref{PeelingEq2}, and Forward Modeling Eq. \ref{ImagingFHDEq}), the majority of the time is spent integrating regions of the model $uv$ plane to form the model visibilities ($\B(\v,\u)$), and then gridding the visibilities with the holographic antenna beams ($\Bv^T(\u,\v)$). 
%Conceptually, the first calculation performs integrations over the $uv$ plane to determine the model visibilities, and the second grids the visibilities back onto the holographic dirty $uv$ plane. 
Instead of repeating these operations for every model component and visibility, we combine them into a Holographic Mapping function. Conceptually the Holographic Mapping function $\H(\u,\u)$ records \textit{\textbf{how}} a $uv$ location in the model is mapped to the holographic $uv$ map, independent of the sky model used and including all of the baseline sampling and direction-dependent beam effects. Effectively we pre-compute the process of forming and gridding visibilities and use this repeatedly as we refine the sky model. The Holographic Mapping function allows us to quickly convert a $uv$ model into an accurate estimate of the holographic $uv$ map as seen by the instrument, including all of the direction-dependent array beam, antenna beam, and polarization effects. The resulting Fast Holographic Deconvolution algorithm is:
\beqa
\gamma\,{\rm MaxCentroid}\Bigg\{ \left[\Ft(\sky,\u)\Bv^T(\u,\v)\v_d\right]_1 - \hspace{1 cm} \nonumber \\ 
\Ft(\sky,\u)\H(\u,\u) \sum_m \F(\u,\sky)\I(\sky_{m})\Bigg\}.
\label{FHDEq}
\eeqa
In words, the source components are used to build a $uv$ model ($\sum_m \F(\u,\sky)\I(\sky_{m})$), the Holographic Mapping function converts this to a model holographic $uv$ map including all of the instrumental and gridding effects, a Fourier transform creates a model holographic dirty image (lower line) which is subtracted from the dirty holographic map of the data [square brackets]. Mathematically FHD is equivalent to the A-projection, peeling and Forward Modeling  deconvolution algorithms,\footnote{Implicit gridding of the model $uv$ plane can introduce small errors far from the field center.}  but is more efficient since the effect of individual direction-dependent antenna gains is pre-computed outside of the deconvolution loop.
% vastly faster. For MWA 32 data the speed advantage is three orders-of-magnitude, and will be greater for future arrays with more antennas.

The speed of Fast Holographic Deconvolution is due to the sparseness of the Holographic Mapping function. Formally $\H(\u,\u)$ is  a very large matrix as it can map any model $uv$ point to any point in the holographic $uv$ map. However, every visibility is the integral of a very limited region of the $uv$ plane---about twice the diameter of an antenna. The gridding function is similarly compact. So in practice a location in the model $uv$ plane is mapped only to nearby locations in the holographic $uv$ plane. This combines with the proximity of nearby frequency and time samples to make the Holographic Mapping matrix  $\H(\u,\u)$ very sparse and amenable to sparse matrix computational methods. Determining the sparse Holographic Mapping function takes about ten times longer than one integration/gridding cycle (Table \ref{comptime} and associated text). But after $\H(\u,\u)$ has been calculated, we can take any model $uv$ contribution and very quickly map it into a holographic dirty map with high precision.

The power of the FHD technique comes from combining image based subtraction and the Holographic Mapping function. This enables us to produce a precision deconvolution method, complete with direction and polarization effects, that is faster than current implementations.% This speed advantage allows us to deconvolve deeper, producing higher quality images.

%We can record how this mapping occurs once, then use it repeatedly for different model sources. Because it records the mapping, all of the direction dependent array beam and antenna gain and polarization effects are included. 

%It is worth noting that Holographic Mapping must be used with image based subtraction. But because the holographic dirty map is lossless, the speed advantage of the holographic mapping allows us to do deconvolution with full direction, polarization, and antenna dependent calibration.

%Mathematically describe the algorithm in the context of other deconvolution approaches. Try to cite them all.

\section{Implementation}
%\subsection{Fast Holographic Deconvolution}
\label{sec:FHD}

%In this section we describe the detailed implementation of our Fast Holographic Deconvolution code. Because we are still developing the basic algorithms, this prototype has been written in the IDL programming language and has not been parallelized. Once algorithm development is complete we intend to translate this code to C or CUDA and parallelize as necessary.

In this section we detail our implementation of  Fast Holographic Deconvolution. The FHD algorithm has three key steps:  creating a holographic dirty image of the data, calculating the Holographic Mapping function $\H(\u,\u)$, and iterating through the deconvolution loop (Equation \ref{FHDEq}). 

\subsection{The Holographic Dirty Image}
\label{sec:FHDgridding}
%\vspace{10 pt}
%\noindent \textbf{The Holographic Dirty Image}

We begin Fast Holographic Deconvolution by constructing a dirty holographic image of the calibrated data for each of the 4 instrumental polarizations. We assume a dual linear reception element (labeled X and Y), and create separate holographic dirty maps from the XX, YY, XY and YX visibilities. The maps are created by gridding each visibility with a kernel based on the individual holographic antenna maps for each polarization, and Fourier transforming the result. It is the holographic gridding kernel that enables the correction of the direction and polarization dependent response of each antenna. Our implementation supports arbitrary antenna kernels and any desired bandwidth synthesis. Since we use image based subtraction (Equation \ref{FHDEq}), we only need to grid the visibilities to the $uv$ plane and FFT once for each instrumental polarization. 

\subsection{The Holographic Mapping Function}
\label{sec:FHDmapfn}
%\vspace{10 pt}
%\noindent \textbf{The Holographic Mapping Function}
%\subsection{The Transfer Function}
%\label{sec:transfunct}

The Holographic Mapping function $\H(\u,\u)$ must be pre-computed for every unique observation and for each instrumental polarization. Once computed, it is stored in the row-indexed sparse storage mode from  \textit{Numerical Recipes in C} and written to disk. Despite being very sparse, this is still typically a large file---a few gigabytes for our example in \S 4. However, the size of the matrix scales with $uv$ coverage rather than the number of baselines or the length of integration, so we believe this approach will remain practical even for future large arrays.
Finally, we note that the choice of pixelization of the $uv$ plane is quite flexible. For example, the dimensions and resolution of the input and output $uv$ planes do not have to match, or one could be a standard grid and the other non-rectilinear such as a $uv$ analog of Healpix.  

\subsection{Deconvolution}
\label{sec:FHDdecon}
%\vspace{10 pt}
%\noindent \textbf{Deconvolution}

Once the Holographic Mapping Function has been built we begin the deconvolution loop described mathematically by Equation \ref{FHDEq}. In words, the steps within each iteration are:

\begin{enumerate}
\item We begin by applying the Holographic Mapping function to the model sky (in $uv$ space) for each instrumental polarization, and take the FFT to create four model dirty holographic images.
\item These are subtracted from the corresponding dirty holographic images of the data. 
\item The residual images are converted to Stokes I, Q, U, and V using the Jones matrices and an average beam model for each instrumental polarization.
\item The Stokes images are median filtered with a small box size to highlight point sources and masked to exclude regions of sky with low instrumental response. (Median filtering is equivalent to down weighting short baselines, but is faster for holographic algorithms.)
\item The positive peak in the filtered Stokes I image is identified. A centroid is fit around the peak and the gain factor $\gamma$ is used to generate a new source component. (Q, U, and V amplitudes of the component may optionally also be fit.)
\item The new component is used to update the $uv$ plane model sky in all four instrumental polarizations.

\end{enumerate}

When deconvolution is finished, we condense all components within a threshold radius of half a pixel to form a list of discrete point source candidates. A robust source list may be constructed by applying standard statistical tests to the list of candidates, such as imposing a minimum signal to noise threshold. Extended sources which have more than a negligible number of close components outside of this threshold may be either left in component form or fit with a model profile.
%These sources are used to generate the final model sky, residual, and restored images (Figure \ref{FigImaging} b-d).

\subsection{Comparison of computational efficiency}
\label{sec:FHDefficiency}
%\vspace{10 pt}
%\noindent \textbf{Comparison of computational efficiency}

Comparing the computational efficiency of very different algorithms is always difficult due to the details of implementation and computer hardware. Instead of counting the formal number of operations, which tends to ignore important effects such as memory bandwidth, we have created matched implementations of both Fast Holographic Deconvolution and A-projection. The IDL programming language was used for both, and no parallelization or optimizations were used (many optimizations work equally well for both). Once we finish algorithm development we intend to translate our FHD implementation to C or CUDA and parallelize as necessary.

%Because we are still developing the basic algorithms, our prototype implementation has been written in the IDL programming language and has not been parallelized. Once algorithm development is complete we intend to translate this code to C or CUDA and parallelize as necessary. However, different styles of deconvolution - whether $A^T$ deconvolution or FHD - will all benefit from being written in a more efficient programming language or run on a faster computer. In this section, we attempt to benchmark the performance of different implementations when run on the same hardware under the same conditions.

%Holographic Deconvolution (HD, equation \ref{ImagingFHDEq}) subtracts the source model directly from the data dirty holographic map, but because it requires constructing model visibilities and gridding them back to the $u,v$ plane it is computationally equivalent to  $A^T$ deconvolution (equation \ref{PeelingEq}). 
The computational cost of deconvolving $N$ components with A-projection can be represented by
\begin{equation}
\tau_{\rm A} =  N \left(  \tau_{\rm model}+\tau_{\rm grid}+\tau_{\rm fit}\right) \label{eq:thd}\\
\end{equation}
where $\tau_{\rm model}$ is the time spent constructing model visibilities ($\Bv^T(\u,\v)$), $\tau_{\rm grid}$ is the time spent gridding the visibilities using the individual direction-dependent antenna gains to the $uv$ plane ($\B(\v,\u)$), and $\tau_{\rm fit}$ is the time spent determining the next model source component. Forward modeling (Equation \ref{ImagingFHDEq}) is computationally similar in cost to A-projection because it requires constructing model visibilities and gridding them back to the $uv$ plane for each model component. 

In Fast Holographic Deconvolution there is an up-front cost for calculating the Holographic Mapping Function,  followed by more efficient determination of each model component. The computational cost of FHD is represented by
%The computation time of each iteration is therefore dominated by the process of constructing the model visibilities ($\Bv^T(\u,\v),  \tau_{\rm model}$) and gridding them using the individual direction-dependent antenna gains ($\B(\v,\u),  \tau_{\rm grid}$), with a relatively small contribution to the time from taking the Fourier transform to the image plane and fitting the brightest source component ($\tau_{\rm fit}$). 
%and Fast Holographic Deconvolution (FHD, equation \ref{FHDEq})
\begin{equation}
\tau_{\rm FHD} = S \left( \tau_{\rm model}+\tau_{\rm grid}\right) +N \left(  \tau_{\rm fit}+\tau_{\rm HMF}\right) \label{eq:tfhd}.  
\end{equation}
The first term is the cost of pre-calculating the HMF, with one iteration of constructing model visibilities and gridding to the $uv$ plane times a computation overhead $S$ associated with building a sparse matrix. The second term is the time required to perform the deconvolution, consisting of the time to apply the HMF $\tau_{\rm HMF}$ and the time to identify the next model component $ \tau_{\rm fit}$ times the number of model components $N$. The speed advantage of FHD comes from moving the costs involved in modeling and gridding visibilities outside of the deconvolution loop. 

Table \ref{comptime} lists the computational cost of each term in Equations \ref{eq:thd} and \ref{eq:tfhd} for our example MWA 32 data (\S \ref{sec:Results}). The total time for deconvolving 20,000 components with the A-projection algorithm ($\tau_{\rm A}$) is more than two orders of magnitude longer than with Fast Holographic Deconvolution ($ \tau_{\rm FHD}$). 

These costs do depend on the details of the instrument. In general $ \tau_{\rm model}$ and $\tau_{\rm grid}$ will scale linearly with the number of visibilities and the size of the gridding kernel, while $\tau_{\rm HMF}$ scales linearly with $uv$ coverage and with the square of the size of the gridding kernel. In practice there are many optimizations that can be applied to both algorithms, many of which will help both approaches. For our IDL implementation of Fast Holographic Deconvolution we have been able to reduce the deconvolution time (2$^{\rm nd}$ term of Equation \ref{eq:tfhd}) to 23 minutes per polarization. Written in a more efficient programming language and parallelized, real-time deconvolution of the example in \S \ref{sec:Results} should be achievable on a high-end desktop computer.

\begin{table}[b] 
\begin{center}   
\caption{\label{comptime} Computational costs}   \small
\hrule
\vspace{6 pt}
%\hline
\begin{tabular}{crc|ccr}
 \multicolumn{2}{c}{Single iteration } &&& \multicolumn{2}{c}{Full deconvolution loop}\\
$\tau_{\rm model}$ & 123 CPU-s &&& $N$ & 20,000 iterations\\
$\tau_{\rm grid}$ & 638 CPU-s &&& $S$ & 9.59\\
$\tau_{\rm fit}$ & 1.66 CPU-s &&& $\tau_{\rm A}$ & 4232 CPU-hours\\
$\tau_{\rm HMF}$ & 3.21 CPU-s &&& $\tau_{\rm FHD}$ & 29.1 CPU-hours
\end{tabular}
\end{center}
\end{table}

\section{Results}
\label{sec:Results}

\begin{figure*}
\begin{center}
\subfigure[Dirty Image]{ \includegraphics[width=3.4in]{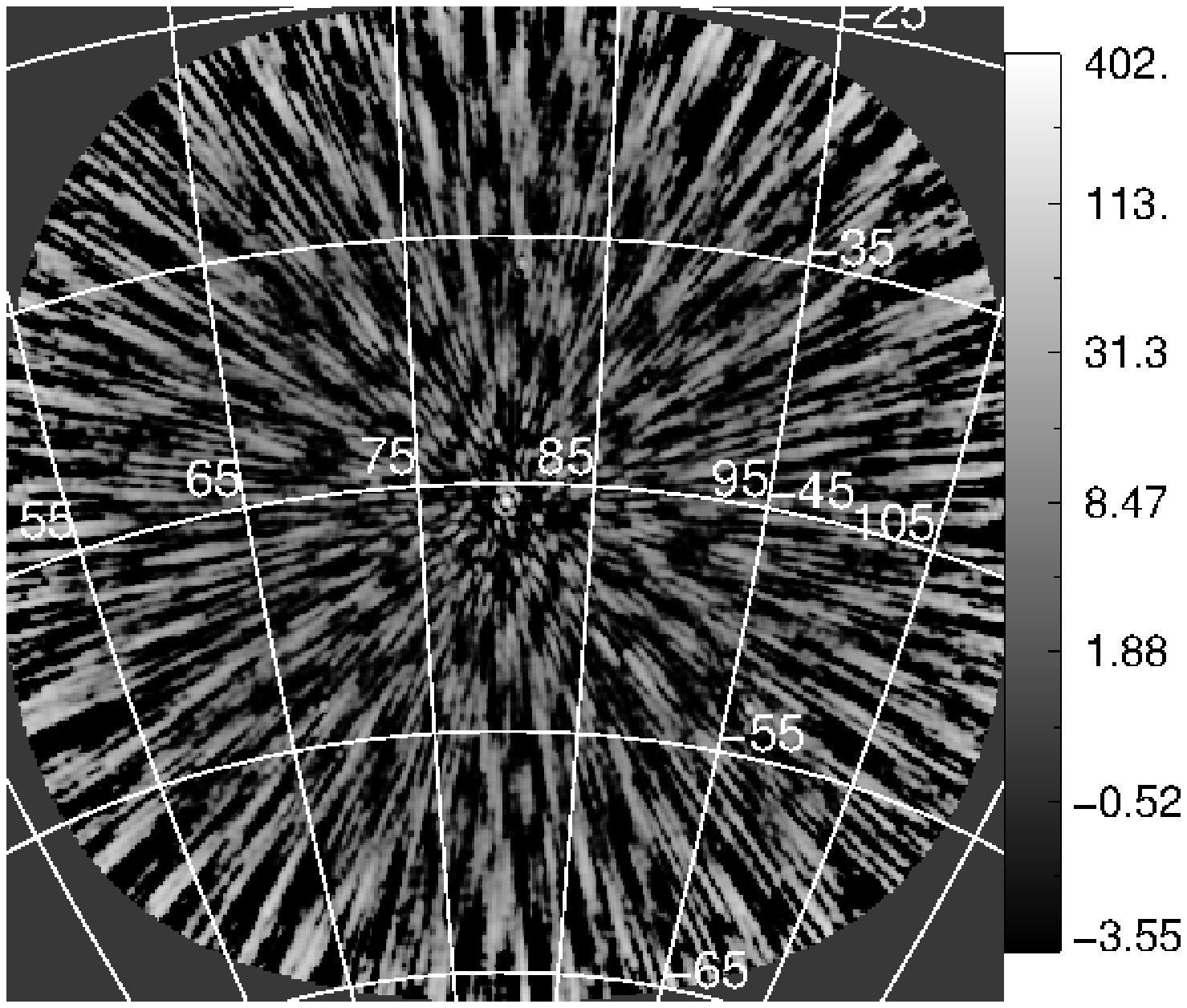}}
\subfigure[Sources]{ \includegraphics[width=3.4in]{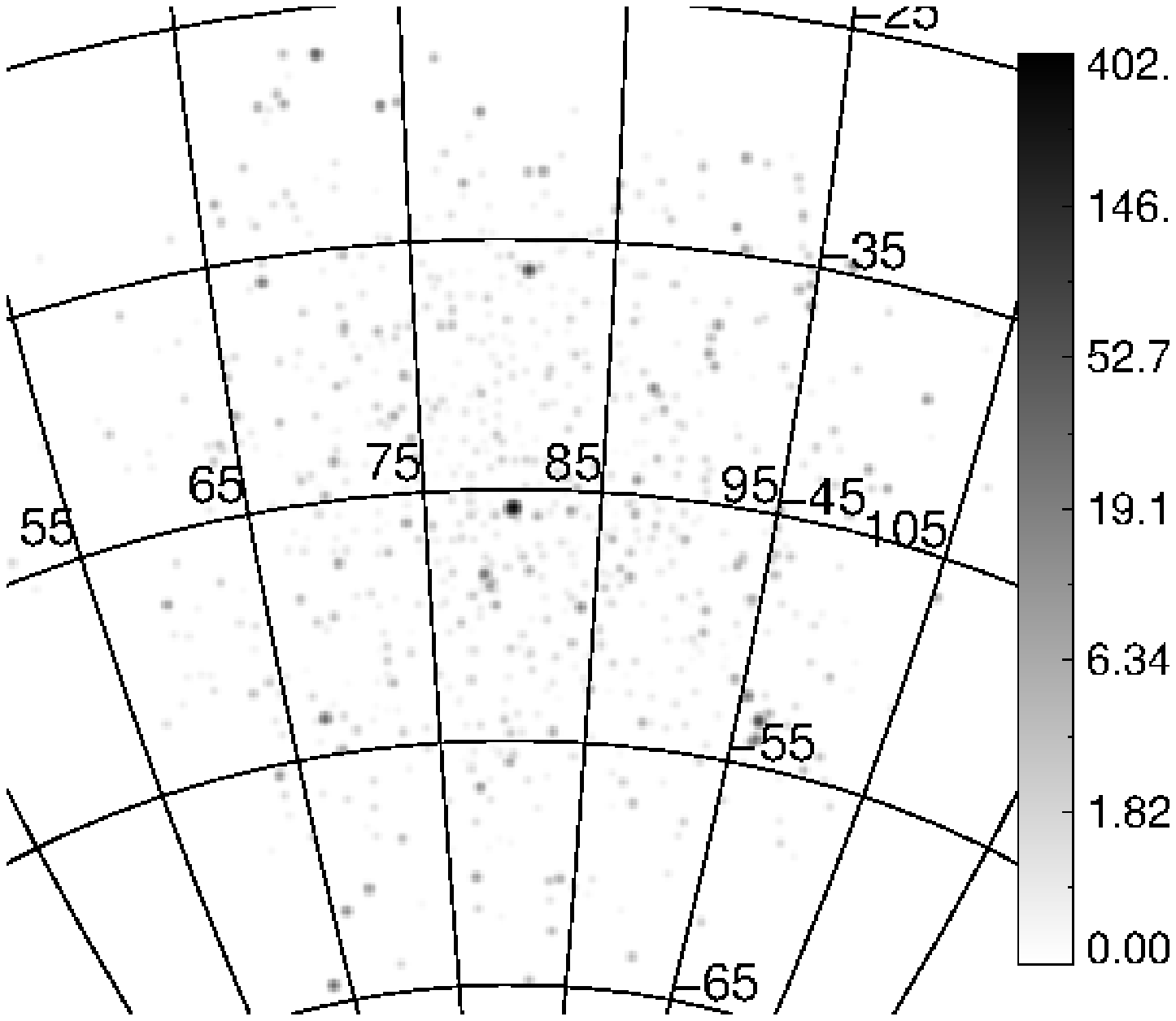}}
\subfigure[Residual Image]{ \includegraphics[width=3.4in]{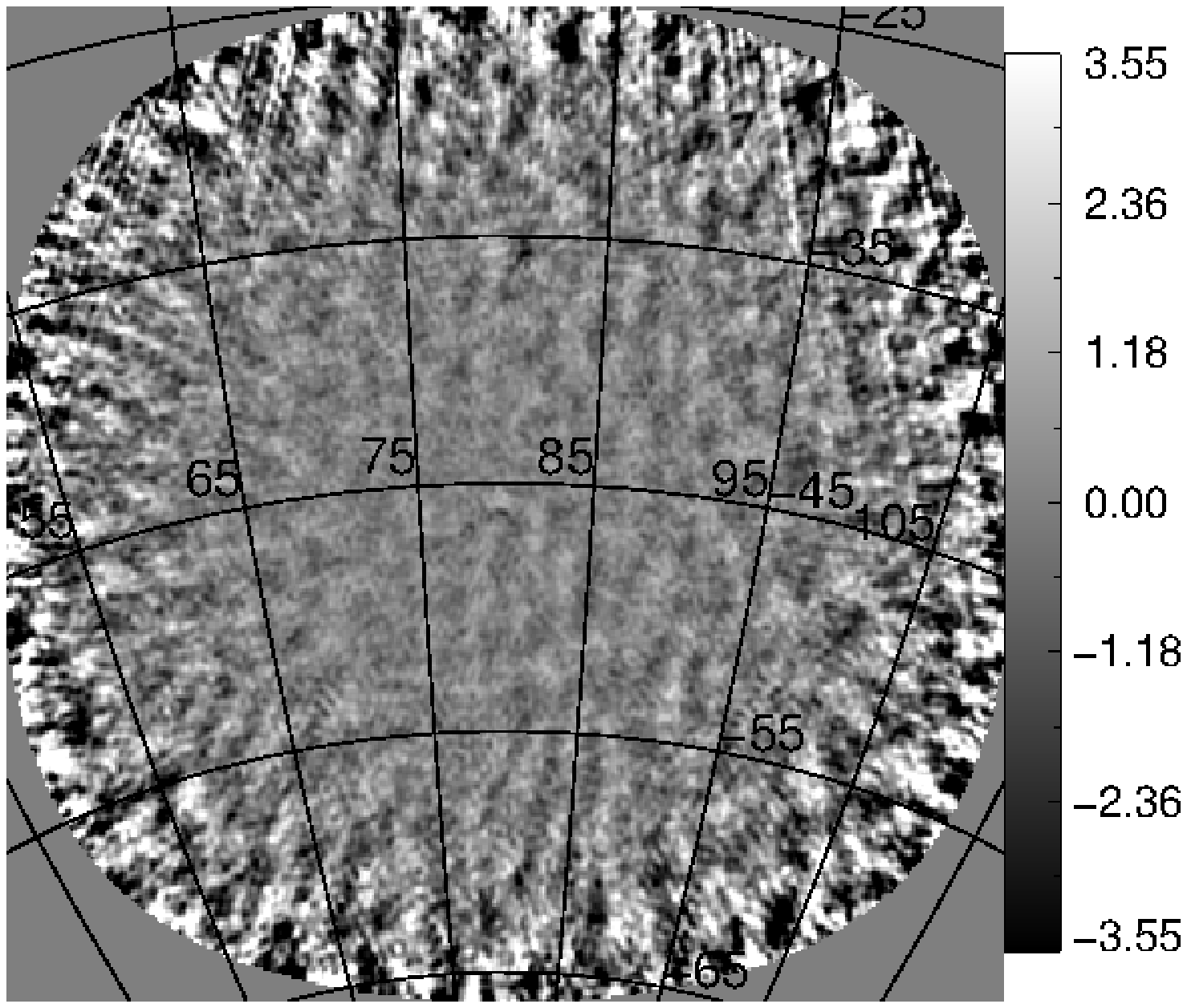}}
\subfigure[Restored Image]{ \includegraphics[width=3.4in]{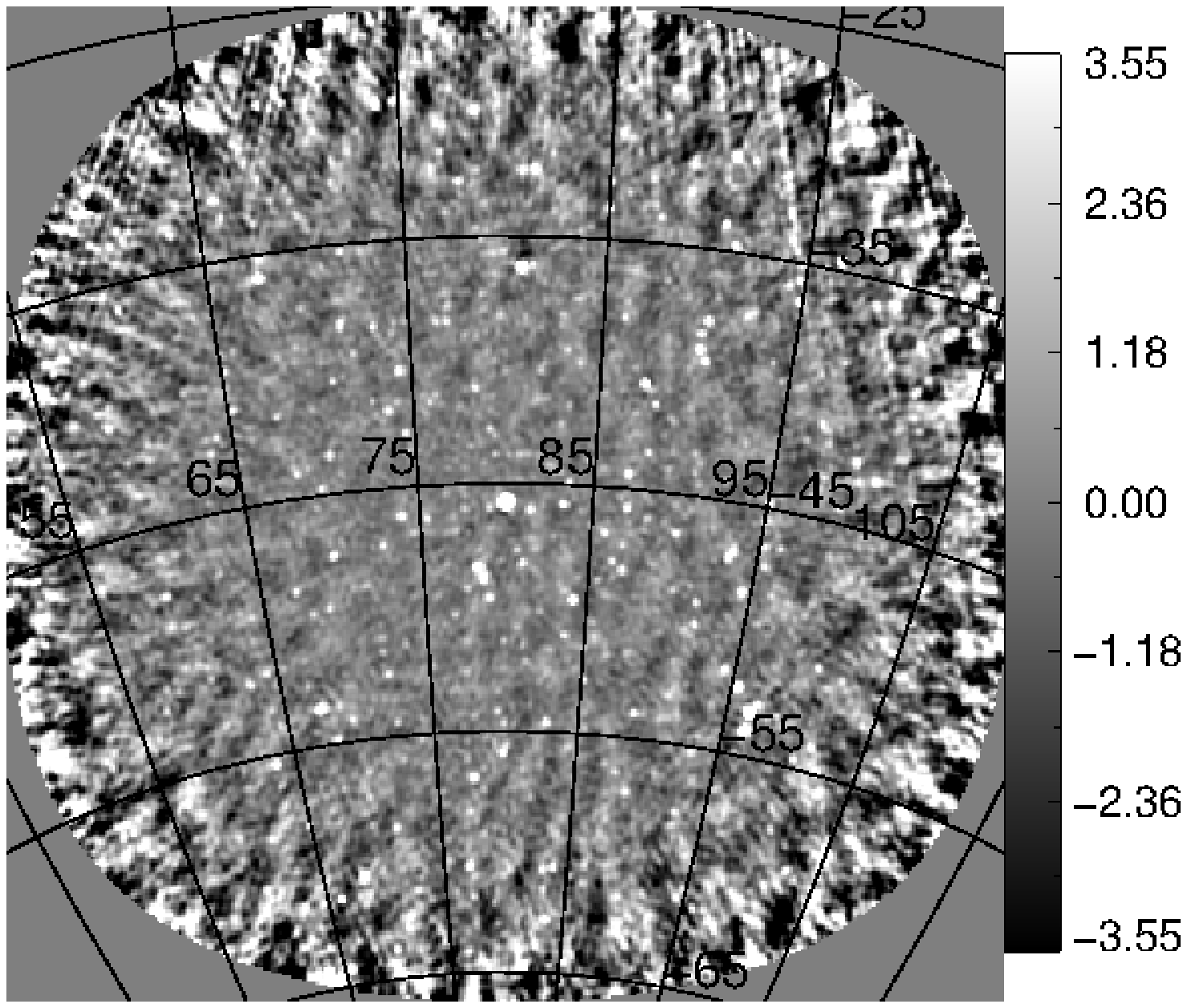}}
\caption{Data from the MWA 32 antenna prototype analyzed with Fast Holographic Deconvolution. The data shown is purposefully a torture test for deconvolution algorithms---a 5 minute snapshot with Pictor~A at the center of the $\sim$ $41^{\circ}\times41^{\circ}$ primary beam FoV, the Orion and Crab Nebulas $40.5^{\circ}$ and $67.9^{\circ}$ to the north of field center (well off the image), and the Galactic plane extending through the antenna side lobes. 
Panel (a) shows the Stokes I dirty image, dominated by Pictor A and the snapshot PSF (array beam) shown on a log stretch to highlight low level detail. Deconvolution reveals 2342 source candidates in this 5 minute snapshot in panel (b) at upper right (log stretch). Panel (c) shows the residual image after deconvolution on an expanded linear scale to highlight the residual emission, and panel (d) shows the reconstructed image (panels b + c) on the same scale. There are false sources in the upper corners of the image due to bright sources outside of the clean region which are not deconvolved. However, we see no residuals from Pictor A or other sources inside the deconvolved region. See the text for a detailed description of the features and Figure \ref{FiglogN} for the $\log N/\log S$ plot and the Stokes Q Image.
}
\label{FigImaging}
\end{center}
\end{figure*}

\begin{figure*}[t]
\begin{center}
\subfigure[Source counts]{ \includegraphics[height=3.1in]{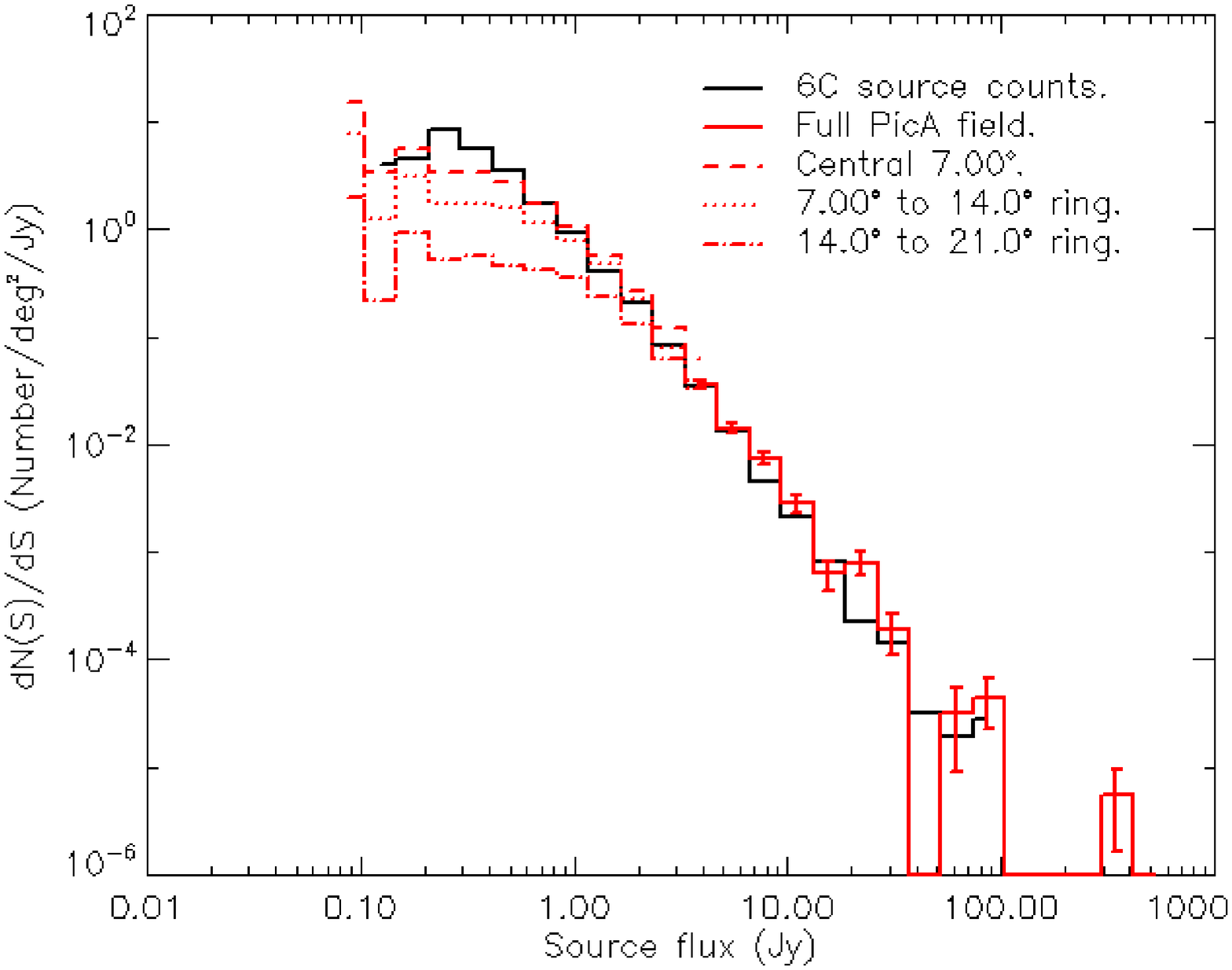}}
\hspace{.1in}
\subfigure[Stokes Q Residual Image]{\includegraphics[width=2.9in]{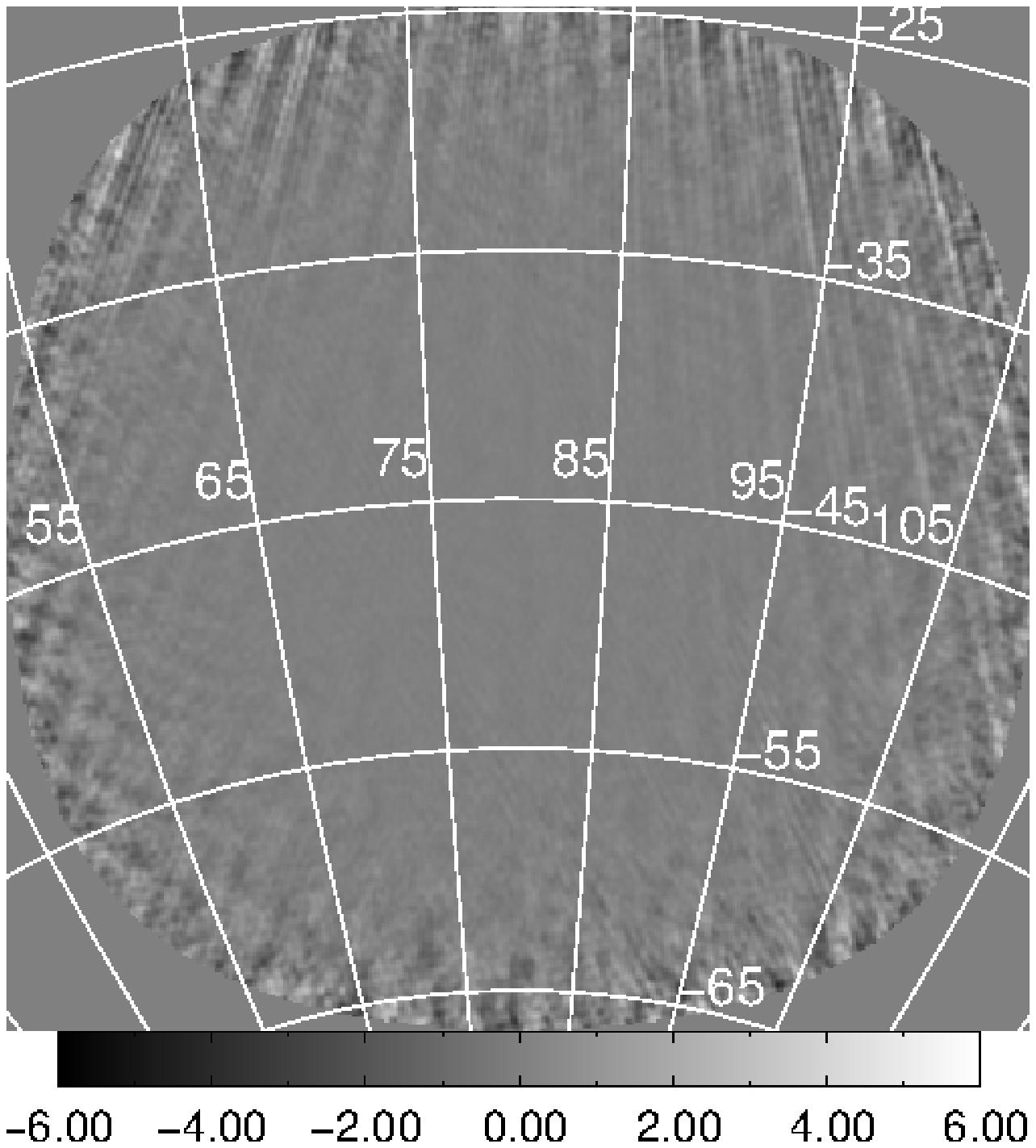}}
\caption{Panel (a) shows the $\log N/\log S$ plot for the sources in Figure \ref{FigImaging}, with error bars to indicate sample variance at high fluxes. We have divided the source statistics below 3~Jy into rings of increasing radius (dashed, dotted, dash-dotted).  Careful investigation of each curve shows the turnover in the completeness increasing with radius as expected. In the central region of the image the sources follow the expected source counts until 600~mJy, which gives us a conservative snapshot dynamic range of $>$800:1. This dynamic range will naturally improve with synthesis rotation (dynamic range becomes hard to measure in longer MWA 32 observations due to the confusion limit). %We believe this is the highest fidelity snapshot produced to date with a HERA\textsuperscript{\ref{fn:hera}} instrument. 
Panel (b) shows the residual Stokes Q image, with fractional polarization of a few percent in the central region ($21^{\circ}$ from antenna bore sight). With upgraded beam models and calibration we expect to achieve percent polarization across the field.
}
\label{FiglogN}
\end{center}
\end{figure*}

In this section we analyze a 5 minute snapshot observation by the 32 antenna MWA prototype \citep{Ord:2010p4920} to demonstrate the effectiveness of Fast Holographic Deconvolution. Detailed descriptions of the full as-built MWA facility can be found in \citep{Tingay2012} and \citep{Lonsdale2009}. This observation is from 139~MHz--169~MHz with Pictor~A at the center of the primary beam field of view (approx.\ $41^{\circ}\times41^{\circ}$ out to the first null), and was taken on September 25$^{\rm{th}}$, 2010 at 5:19~am local time (21:19 UT) on our 14$^{\rm{th}}$ field expedition. The Orion and Crab Nebulas are $40.5^{\circ}$ and $67.9^{\circ}$ to the north of field center in the first and second antenna side lobes, and the Galactic plane extends through the southern side lobes. The results of our FHD algorithm are shown in Figures \ref{FigImaging} and \ref{FiglogN}. No attempt to deconvolve sources in the sidelobes of the array beam was made, so artifacts from the bright Crab nebula are visible in the Northern corners of panels (b)-(d). 

Panel (a) of Figure \ref{FigImaging} shows the holographic dirty image prior to deconvolution, which is dominated by Pictor A and the array beam (PSF). We perform 20k iterations of deconvolution with a clean gain of 15\% using only positive, centroided source components and removing all peaks above 600~mJy in the central region of the residual image. This peak detection threshold rises as sensitivity decreases towards the edge of the field to maintain a constant minimum signal to noise. We then collect all components within 6 arc minutes to identify 2342~source candidates, shown in panel (b) [because each component is independently centroided, individual components of a real point source will differ slightly in location by small fractions of a pixel]. The residual image is shown in panel (c) on an expanded linear scale to highlight the fluctuations (no smoothing or filtering has been applied to any of these images). There are very few features in the residual image, the residuals are highly Gaussian with noise of $\sigma$=230~mJy in the center, and there is no evidence of the Pictor A array beam. As expected, the noise increases towards the beam edge where the sensitivity is down by a factor of 20. The flux density of Pictor~A has been scaled to match the catalog value of 402~Jy to set an approximate overall scale.

The source counts are shown in panel (a) of Figure \ref{FiglogN}.  The black solid line shows the counts from the $6^{\rm th}$ Cambridge survey at 151~MHz in the Northern hemisphere \citep{Hales1988}, overlaid by our source counts in red. Because we deconvolve sources out to near the first null of the tile beam, our sensitivity naturally falls towards the field edge. At the faint end of the distribution (below 3~Jy) we divide our source counts into rings of nearly equal sensitivity (dashed lines). At the bright end (above 3~Jy) we use the full field as the uncertainty becomes dominated by the low source counts. There appears to be good correspondence between the source counts from FHD and the expected counts, with the expected roll-offs in sensitivity in the outer edges of the field. We observe excellent agreement of all bright source candidates with the Molonglo Reference Catalog (MRC) of radio sources \citep{Large1981}, though we defer a thorough catalog comparison to future work with deeper integrations.
%The theoretical point source sensitivity in the center of the field is 140~mJy.

HERA\footnote{Hydrogen Epoch of Reionization Arrays \href{http://reionization.org/}{(reionization.org)}\label{fn:hera}} instruments such as the 32 antenna MWA prototype quickly reach the confusion limit (expected to be $\sim$300~mJy). This makes measuring the dynamic range of a deconvolution algorithm problematic using common measures such as the RMS in blank areas of the field. Within $7^{\circ}$ of the field center, the source counts appear to follow the 6C catalog to 600~mJy. This implies a minimum snapshot dynamic range for Fast Holographic Deconvolution of 800:1 in this data set. This dynamic range will naturally improve with synthesis rotation. The polarization performance is represented by the residual Stokes Q image in Figure \ref{FiglogN}b, with a fractional polarization of a few percent in the central region. With improved beam models and calibration we expect to achieve percent polarization residuals across the field.

\section{Conclusion}
\label{sec:Conc}

A-projection and forward modeling deconvolution algorithms that use direction-dependent gain and polarization are very precise, but are much slower than the traditional CLEAN approach. In this paper we have introduced Fast Holographic Deconvolution (FHD), which uses the Holographic Mapping function to improve the speed of holographic deconvolution without sacrificing precision. We have demonstrated the power of the FHD algorithm using a 5 minute snapshot of data from the 32 tile MWA prototype.

In future work we plan to enhance the capabilities of our Fast Holographic Deconvolution algorithm; including parallelizing and optimizing our implementation, deconvolving diffuse sources, and implementing w-projection to enable long widefield integrations (\citealt{Cornwell:2008p4785}; using holographic formulation in \citealt{Morales:2009p4730}). Because w-projection decreases the sparseness of the Holographic Mapping function (larger gridding kernels) it may be more efficient to do a joint deconvolution of a few intermediate length observations than one long track. We will also explore ways of using sparse matrix parallelization in addition to the more traditional frequency or time parallelizations. Finally, Fast Holographic Deconvolution lends itself naturally to modeling diffuse sources such as the galaxy, and we plan to explore multi-scale or other principle component deconvolution methods.

\section*{Acknowledgments}
Support for this work came from the U.S. National Science Foundation (grants AST CAREER-0847753, AST-1003314, AST-0457585, AST-1008353, AST-0908884, PHY-0835713), the Australian Research Council (grants LE0775621 and LE0882938), the U.S. Air Force Office of Scientific Research (grant FA9550-0510247), the Smithsonian Astrophysical Observatory, the MIT School of Science, the Raman Research Institute, the Australian National University, the iVEC Petabyte Data Store, the Initiative in Innovative Computing and NVIDIA sponsored Center for Excellence at Harvard, and the International Centre for Radio Astronomy Research, a Joint Venture of Curtin University of Technology and The University of Western Australia, funded by the Western Australian State government. 

We acknowledge the Wajarri Yamatji people as the traditional owners of the Observatory site.

\bibliographystyle{apj}
\bibliography{morales,Imaging1_extra}

\end{document}